\magnification=1100
\voffset=.5truecm
\hoffset=-.5truecm
\vsize=24truecm
\hsize=17truecm
%\baselineskip=.9truecm plus .01truecm minus .01truecm
\parskip=7pt
\null

\font\tbf=cmbx10 scaled \magstep2
\hyphenation {Schwarz-schild}
\hyphenation {Abra-mo-wicz}
\font\tbmi=cmmib10
\def\blambda{\hbox{\tbmi\char'025}}

\def\btimes{\hbox{\tbs\char'002}}
\def\bnabla{\hbox{\tbs\char'162}}
\font\tenmib=cmmib10 \textfont"E=\tenmib
\font\tenbsy=cmbsy10 \textfont"F=\tenbsy

\mathchardef\blambda="0E15
\mathchardef\bnabla="0F72
\mathchardef\btimes="2F02 

\def\ref{\par\noindent\hangindent=1 truecm}
\def\simless{\mathbin{\lower 3pt\hbox
   {$\rlap{\raise 5pt\hbox{$\char'074$}}\mathchar"7218$}}}
\def\simgreat{\mathbin{\lower 3pt\hbox
   {$\rlap{\raise 5pt\hbox{$\char'076$}}\mathchar"7218$}}}

\null

\vskip 3truecm

\centerline {\tbf Epicyclic orbital oscillations}
\centerline {\tbf in Newton's and Einstein's dynamics}
\vskip 0.7truecm

\centerline {Marek A. Abramowicz$^1$ and W{\l}odek Klu{\'z}niak$^2$}

\centerline {Institut d'Astrophysique de Paris, 98bis Boulevard Arago,
75014 Paris, France}

\vskip 0.4truecm

\centerline {$^1$ Department of Astrophysics, Chalmers University, 412-96
G{\"o}teborg, Sweden}

\centerline {$^2$ Institute of Astronomy, Zielona G{\'o}ra University,
Lubuska 2, 65-365 Zielona G{\'o}ra, Poland}

\vskip 0.3truecm

\centerline {marek@fy.chalmers.se, wlodek@camk.edu.pl}

\vskip 0.8truecm
\centerline {\bf Abstract}
\vskip 0.2truecm

\noindent We apply Feynman's principle, ``The same equations have the same
solutions'', to Kepler's problem and show that Newton's dynamics in a
properly curved 3-D space is {\it identical} with that described by
Einstein's theory in the 3-D optical geometry of Schwarzschild's
spacetime. For this reason, rather unexpectedly, Newton's formulae for
Kepler's problem, in the case of nearly circular motion in a static,
spherically spherical gravitational potential {\it accurately} describe
strong field general relativistic effects, in particular vanishing of the
radial epicyclic frequency at $r = r_{\rm ms}$.

\vskip 0.8truecm
\leftline {\bf 1. Introduction: observed QPOs as a test of strong gravity}
\vskip 0.2truecm

\noindent The standard accretion disk theory (Shakura and Sunyaev, 1973)
assumes that matter in accretion disks 
around black holes and neutron stars moves on nearly circular, nearly
Keplerian (i.e., nearly geodesic) orbits. Theory predicts several strong
field general relativistic effects that should follow directly from this
assumption, but none of them has so far been clearly detected. Most
recently, we have found a new effect of this type (Klu{\'z}niak and
Abramowicz, 2000, 2001, 2002).
It concerns the QPOs, quasi periodic oscillations
with kilohertz frequencies, observed as variations in the X-ray luminosity
of accreting neutron stars and black holes. QPOs often occur in coupled
pairs manifested as characteristic double peaks in the variability power
spectra. According to us, observed frequencies at the double peaked QPOs
are directly related to orbital epicyclic frequencies, vertical and
radial, which are in a $3:2$ parametric resonance. The resonance occurs in
super strong gravity, just several gravitational radii outside the central
black hole or neutron star, at a precisely determined radius. Knowing
that, one may sharply constrain the global parameters of the source, as it
was done, for example, for the Kerr angular momentum parameter in the
black hole ``candidate'' GRO J1655-40 (Abramowicz and Klu{\'z}niak, 2001)
that shows two QPOs at $450~$Hz and $300~$Hz (Strohmayer, 2001).

\noindent In strong Einstein's gravity the radial epicyclic frequency
$\Omega_r$ is smaller than the Keplerian orbital frequency $\Omega_K$, and
at the radius of the marginally stable orbit $\Omega_r = 0$. In weak
Newton's gravity one has $\Omega_r = \Omega_K \not = 0$. These very
different behaviours are attributed by many authors to a ``non-linearity''
of Einstein's gravity. We discuss here a simpler and more proper
explanation: the difference is due {\it only} to the curvature of the
three dimensional space.

\noindent Our explanation consists of three steps. First, we recall the
relevant Einstein's equations. We write them in a particular form,
consistent with the optical geometry of space. Second, we re-derive
well-known, standard Newton's equations using a particular notation, and
prove that they are identical in form with the corresponding Einstein's
equations in optical geometry. Third, we apply Feynman's principle, ``The
same equations have the same solutions'' (e.g. Feynman et al., 1989), and
derive the formula for the epicyclic radial frequency that is valid in
both Einstein's and Newton's gravity.

\vskip 0.6truecm
\leftline {\bf 2. Optical geometry in Schwarzschild spacetime}
\vskip 0.2truecm

\noindent The general static, spherically symmetric metric can be written
in a particular form,

$$ ds^2 = e^{2 \Psi} 
\left \{ 
- c^2 dt^2 + 
\left [ 
dr^2_* + {\tilde r}^2
\left ( d \vartheta^2 + \sin^2 \vartheta d\varphi^2 
\right ) 
\right ]
\right \}.
\eqno (2.1)
$$

\noindent In the specific case of Schwarzschild geometry one has,

$$ e^{2 \Psi} = 1 - {r_G \over r},~~~ dr_* = e^{- 2 \Psi} dr,~~~{\tilde
r}^2 = e^{-2 \Psi} r^2,
\eqno (2.2)
$$

\noindent where $r_G = 2GM/c^2$ is the gravitational radius of the central
body with the mass $M$. The 3-D metric of {\it optical geometry} was
introduced by Abramowicz, Carter and Lasota, (1988). It corresponds to the
part of (2.1) in square brackets,

$$
ds^2_{\rm optical} = dr^2_* + {\tilde r}^2 \left (
     d \vartheta^2 + \sin^2 \vartheta d\varphi^2 \right ), 
\eqno (2.3) 
$$

\noindent In terms of optical geometry the Schwarzschild quantities (2.2)
have a very clear geometrical meaning. Obviously, $r_*$ is the {\it
geodesic} radius, and ${\tilde r}$ is the {\it circumferential} radius of
the nested, concentric spheres $r =$ const that generate the spherical
symmetry of the metrics (2.1) and (2.3). 

\noindent Let us introduce the covariant derivative operator $\nabla^i$ in
the optical geometry (2.3). From the $R_{tt} = 0$ component of Einstein's
field equations one easily derives,

$$
\nabla^2 \Phi = \nabla^i \nabla_i \Phi = 0,~~~\Phi = {c^2 \over 2}
\left ( e^{2 \Psi} - 1 \right )
\eqno (2.4)
$$

\noindent This optical geometry equation for $\Phi$ is {\it linear} and
identical with Laplace's equation that gravitational potential obeys in
Newton's theory. This, together with the asymptotic behaviour of $\Phi$
for large $r$, namely $\Phi = - GM/r$, suggests that $\Phi$ should be
called the gravitational potential. We shall see later more reasons to do
so.
     
\noindent Equation of geodesic motion along a great circle on a $r =$
const sphere, and with a constant speed $v = c \beta \gamma$, takes the
form, valid both in the optical geometry, and in the full 4-D spacetime
(Abramowicz, Carter and Lasota, 1988),

$$ a^i = c^2 \nabla^i \Psi - {v^2 \over {\cal R}}\lambda^i = 0. 
\eqno (2.5) 
$$
   
\noindent Here $a^i$ is the four-acceleration,
 $\gamma = (1 -\beta^2)^{-1/2}$ 
is the gamma Lorentz factor, $\lambda^i=\nabla^i r_*$ is the first
normal to the circle 
(orthogonal vector, of unit length in the metric of eq. [2.3]), and 
${\cal R}$ is the {\it
curvature} radius of the sphere. (For simplicity, one may consider a
particular great circle, located at the {\it equatorial plane}, 
$\vartheta= \pi/2$, but our arguments are valid in a general case).

\noindent The conserved angular momentum equals (Abramowicz, Carter and
Lasota, 1988),

$$ {\cal L} = v {\tilde r} e^{\Psi}. \eqno (2.6) $$

\noindent We use (2.6) to write the final formula,

$$ a^i = e^{-2 \Psi} \left [ \nabla^i \Phi - {{\cal L}^2 \over {{\cal R}
{\tilde r}}}\lambda^i \right ] =0. 
\eqno (2.7) 
$$
 
\noindent In these calculations one may use a convenient relation,

$$ {d {\tilde r} \over dr_*} = {{\tilde r} \over {\cal R}}. 
\eqno (2.8) 
$$

\noindent For light trajectories $v = \infty$, and from (2.5) it
follows that also ${\cal R} = \infty$, which means that light may go round
a circular trajectory if and only if this trajectory is a geodesic circle
in optical geometry --- a conclusion that follows also from Fermat's
principle: {\it light rays coincide with geodesic trajectories in optical
geometry} (Abramowicz, Carter and Lasota, 1988; Abramowicz, 1994).

\noindent Equations (2.4) and (2.7) that we have recalled from Einstein's
theory will be compared with corresponding Newton's equation that we
derive next.

\vskip 0.5truecm
\leftline {\bf 3. Newton's dynamics in a curved 3-D space}
\vskip 0.2truecm

\noindent Newton's dynamics is usually considered in 3-D Euclidean space,
but its generalization to a curved space is trivial. Indeed, the only
issue that is important in the present context is a careful distinction
between the three radii of a sphere: geodesic radius $r_*$,
circumferential radius ${\tilde r}$, and curvature radius ${\cal R}$. In
Newton's theory, with Euclidean geometry assumed, one has $r_* = {\tilde
r} = {\cal R}$, but assuming Euclidean geometry is not necessary in
Newton's dynamics, and one could easily distinguish the three different
radii in all calculations.
  
\noindent Indeed, let us consider a curve in space defined by,

$$ x^i = x^i(s), \eqno (3.1) $$

\noindent where $x^i$ are coordinates in a coordinate system, $s$ is the
length along the curve, and Latin indices run through $1,~2,~3$. The
velocity $v^i$ is defined by,

$$ v^i = {dx^i \over {dt}}, \eqno (3.2) $$

\noindent where $t$ is the absolute time. From this definition it follows that

$$ v^i = {dx^i \over {dt}} = {dx^i \over {ds}} {ds \over dt} = \tau^i v, \eqno
(3.3) $$

\noindent where $\tau^i = dx^i/ds$ is a unit tangent vector to the curve
(3.1), and $v = ds/dt$ is the speed along the curve.

\noindent Acceleration is defined as

$$ a^i = {dv^i \over {dt}}, \eqno (3.4) $$
 
\noindent and from this definition it follows

$$ a^i = {dv^i \over {dt}} = {d\over dt}(\tau^i v) = {ds \over dt} {d\over
ds}(\tau^i v)  = v^2 {d\tau^i \over ds} + v \tau^i {dv \over ds} = 
-{v^2 \over {\cal R}}\lambda^i + \tau^i {d{\cal E}_K \over ds}, \eqno 
(3.5) $$

\noindent where $-\lambda^i$ is the first normal to the curve (3.1), ${\cal
R}$ is the curvature radius of the curve (3.1) and ${\cal E}_K$ is the
kinetic energy per unit mass. Because ${\cal E}_K$ is obviously constant
for a circular motion with a constant speed, we may write that for such
motion,

$$ a^i = -{v^2 \over
{\cal R}}\lambda^i.  \eqno (3.6) $$
 
\noindent Newtonian dynamics is based on the second law,

$$ F^i = m a^i, \eqno (3.7) $$

\noindent where $F^i$ is the applied force and $m$ the mass. Because we are
interested here in a ``Keplerian" motion, with

$$ F^i = -m \nabla^i \Phi \eqno (3.8) $$

\noindent being the gravitational force (equal to the gradient $\nabla^i$
of gravitational potential $\Phi$) the second law takes the form,

$$ \nabla^i \Phi = {v^2 \over {\cal R}}\lambda^i. \eqno (3.9) $$

\noindent Using Newton's formula for angular momentum,

$$ {\cal L} = v {\tilde r}, \eqno (3.10), $$

\noindent we write finally,

$$ \nabla^i \Phi = {{\cal L}^2 \over {{\cal R} {\tilde r}}}\lambda^i. 
\eqno (3.11) 
$$
 
\noindent This is identical with Einstein's equation (2.7). 

\noindent The gravitational potential obeys Laplace's equation,

$$
\nabla^i \nabla_i \Phi = 0,
\eqno (3.12)
$$

\noindent which is identical with Einstein's equation (2.4). We thus have
completed the second step, showing that for circular motion in spherical
potential, Einstein's equations in optical geometry, and Newton's
equations are the same.

\vskip 0.5truecm
\leftline {\bf 4. The same equations have the same solutions}
\vskip 0.2truecm

\noindent Einstein's and Newton's equations for Kepler's circular motion
in a spherical potential have the same form,

$$ \nabla^i \Phi = {{\cal L}^2 \over {{\cal R} {\tilde r}}}\lambda^i,
\eqno (4.1)
$$

$$
\nabla^i \nabla_i \Phi = 0,
\eqno (4.2)
$$

\noindent and the physical and geometrical meaning of all quantities
appearing in them is the same.

\noindent Let us integrate Laplace's equation (4.2) in the volume ${\cal
V}$ between two equipotential surfaces, ${\cal S}_1$, defined by $\Phi_1 =
$ const, and ${\cal S}_2$, defined by $\Phi_2 = $ const, and use the Gauss
theorem,

$$ \int_{\cal V} \nabla^i (\nabla_i \Phi) dV = \int_{{\cal S}_1 + {\cal
S}_2} (\nabla_i \Phi) \lambda^i_* dS = \int_{{\cal S}_1} (\nabla_i \Phi)
\lambda^i dS - \int_{{\cal S}_2} (\nabla_i \Phi) \lambda^i dS = 0, \eqno
(4.3) $$

\noindent where $\lambda^i_*$ is the vector orthogonal to the surface
${\cal S}$ --- obviously, $\lambda^i_* = \lambda^i$ at ${\cal S}_1$ and
$\lambda^i_* = - \lambda^i$ at ${\cal S}_2$. Let us fix a position of
${\cal S}_1$, and denote

$$ \int_{{\cal S}_1} (\nabla_i \Phi) \lambda^i dS = C. \eqno (4.4) $$  

\noindent If we now change the position of ${\cal S}_2$ then, because
(4.3) holds independently of the position of ${\cal S}_2$, and $(\nabla_i
\Phi )\lambda^i$ is constant on each equipotential surface, one must
conclude that

$$ (\nabla_i \Phi) \lambda^i \int_{{\cal S}}dS = C = {\rm const}, \eqno
(4.5) $$

\noindent for {\it any} equipotential surface ${\cal S}$. Taking into
account that ${\tilde r}$ is the circumferential radius of the sphere, and
therefore

$$ \int_{{\cal S}}dS = 4\pi {\tilde r}^2, \eqno (4.6) $$ 

\noindent one concludes that

$$(\nabla_i \Phi) \lambda^i = {C \over 4\pi{\tilde r}^2} = {GM \over {\tilde
r}^2}, \eqno (4.7)$$

\noindent with $C=4\pi GM$ following from the asymptotic behaviour at
${\tilde r} \rightarrow \infty$.

\par \noindent Thus, finally, the second law takes the form,

$$ {GM \over {\tilde r}^2} = {{\cal L}^2 \over {{\tilde r}^2 {\cal R}}}, 
\eqno (4.8) 
$$

\noindent and from this we derive the formula for the Keplerian angular
momentum distribution,

$$ {\cal L}^2 = GM{\cal R}.  \eqno (4.9) $$

\noindent The above formula allows a novel and interesting
interpretation of the Keplerian angular momentum, as the geometrical
mean of the gravitational radius of the gravitating center $r_G =
2GM/c^2$, and of the curvature radius ${\cal R}$ of the particle
trajectory,

$$ {\cal L} = c \sqrt { 2 r_G {\cal R}}.  \eqno (4.10) $$

\noindent One knows that the circular photon trajectory is located at $r =
(3/2) r_G$. This means that ${\cal R} = \infty$ there. But ${\cal R} =
\infty$ also for $r = \infty$. This means, that somewhere in the range
$[(3/2) r_G, \infty]$ the curvature radius ${\cal R}$ must have a minimum.
Because ${\cal L}^2 = GM{\cal R}$, the angular momentum ${\cal L}$ also
has a minimum at the same radius. It must be the radius $r = r_{ms} =
3r_G$ of the mariginally stable orbit.

\noindent To see this in a more quantitative way, let us now consider a
small radial perturbation of Einstein's equation (2.7), which is
equivalent to (4.1). We assume that a particle with a fixed angular
momentum ${\cal L}$ is displaced from its original orbit at $r_*$ to a new
one, at $r_* + \delta r_*$. The perturbation introduces an unbalanced
force and radial acceleration,

$$ 
e^{2 \Psi} \lambda_i a^i = e^{4 \Psi} {{d^2} \over d s^2} \delta r_* =
e^{4 \Psi} A^2 {{d^2} \over d t^2} \delta r_* = - {1 \over {{\tilde r}^2
{\cal R}}}{{d {\cal L}^2} \over dr_*} {\delta r_*}, 
\eqno (4.11) 
$$

\noindent where $A$ denotes the total redshift factor, connected to
the previously introduced Lorentz gamma factor by

$$
A^2 = \gamma^2 e^{- 2 \Psi} =\left ( 1 - {3 r_G\over 2 r} \right )^{-1}.
\eqno (4.12)
$$   

\noindent From (4.11) we derive, finally, 

$$
{\ddot {\delta r_*}} + 
\Omega_K^2 
\left({{d {\cal R}} \over dr_*} {{\tilde r}^2\over {\cal R}^2}\right)
{\delta r_*} = 0.
\eqno (4.13)
$$
 
\noindent Here, each dot denotes a differentiation with respect to
time of the observer at infinity, and

$$
\Omega_K = {\cal L}e^{-\Psi}\tilde r^{-2}  A^{-1}= 
 \left({GM \over r^{3}}\right)^{1/2}
\eqno (4.14)
$$ 

\noindent is the Keplerian orbital frequency, also observed at infinity.  

\noindent It is quite remarkable that equation (4.13) has identical form
in Einstein's and Newton's theories --- the well-known equation for a
harmonic oscillator. Its eigenfrequency,

$$
\Omega_r^2 = 
\left({{d {\cal R}} \over dr_*} {{\tilde r}^2 \over{\cal R}^2}\right)
\Omega_K^2 
\eqno (4.15)
$$
 
\noindent is obviously equal to the epicyclic frequency of small radial
oscillations. When $\Omega_r^2 > 0$ the radial epicyclic oscillations are
stable, and when $\Omega_r^2 < 0$, they are unstable. Thus, mariginal
stability occurs at the radius where $d{\cal L}/dr = 0 = d{\cal R}/dr$
which, in Schwarzschild geometry is at $r = r_{ms}$.

\noindent In Newton's case the geometry is flat (Euclidean) and the
geodesic, circumferential and curvature radii are equal,

$$ r_* = {\tilde r} = {\cal R}.
\eqno (4.16)
$$

\noindent From these expression and (4.13) one derives,

$$
\Omega_r^2 = \Omega_K^2~~~{\rm in~Newton's~theory}. 
\eqno (4.17)
$$

\noindent This means that the radial epicyclic frequency equals to the
Keplerian orbital frequency, which is why Newton's orbits are closed
ellipses.

\noindent In Einstein's gravity, the geometry is curved and the
geodesic, circumferential and curvature radii are all different,

$$
r_* = \int \left ( 1 - {r_G \over r}\right )^{-1} dr,~~~{\tilde r} =
r \left ( 1 - {r_G \over r}\right )^{-1/2},~~~
{\cal R} = r \left ( 1 - {3 r_G \over 2 r}\right )^{-1}.
\eqno (4.18)
$$

\noindent From these expressions and (4.13) one recovers the well-known
formula, to our best knowledge first derived by Kato and Fukue (1980),

$$
\Omega_r^2 = \Omega_K^2 \left ( 1 - {3r_G \over r}\right ), ~~~{\rm
in~Einstein's~theory}.
\eqno (4.19)
$$

\noindent This completes our point: Einstein's and Newton's formulae (4.1)
and (4.2) that describe dynamics in static, spherically symmetric gravity
are the same in both theories (and in both theories {\it linear}). For
this reason, the formula for radial epicyclic motion (4.13), derived
directly from these equations, is also the same in both theories. However,
this formula depends on geodesic, circumferential and curvature radii of
circular trajectories. In the flat, Euclidean, geometry of Newton's space
these radii are equal, and in the curved geometry of Einstein's
space they are not. This geometrical difference alone, and {\it not} an
often mentioned (but not present in this case) ``non-linearity'' of
Einstein's equations, is the reason for the distinctively different
Newton's and Einstein's predictions for the physical behaviour of small
radial oscillations around Keplerian circular orbits in static spherically
symmetric gravity.

\noindent The same conclusion was previously reached by Abramowicz, Lanza,
Miller and Sonego (1997) for weak gravity, who considered the perihelion
of Mercury advance according to Newton's gravity in a properly curved
space. It is rather surprising that the conclusion holds also for
arbitrarily strong static spherically symmetric gravity.

\vskip 0.5truecm
\leftline {\bf References}
\vskip 0.2truecm

\noindent Abramowicz M.A., Carter B., and Lasota J.-P., 1988, Gen. Rel.
Grav., {\bf 20}, 1173

\noindent Abramowicz M.A., 1994, Sci. Amer., March issue

\noindent Abramowicz M.A., Lanza A., Miller J.C., and Sonego S., 1997,
Gen. Rel. Grav., {\bf 29}, 1997

\noindent Abramowicz M.A., and Klu{\'z}niak W., 2001,
Astron. Astrophys., {\bf 374}, L19  

\noindent R. P. Feynman, R. B. Leighton, and M. Sands, The Feynman
   Lectures on Physics, 1989, Addison-Wesley Pub. Co., Section 12-1

\noindent Kato S., and Fukue J., 1980, Publ. Astron. Soc. Japan, {\bf 32},
377

\noindent Klu{\'z}niak W., and Abramowicz M.A., 2000, submitted to Phys.
Rev. Lett.,  astro-ph/0105057 

\noindent Klu{\'z}niak W., and Abramowicz M.A., 2001, Acta Phys. Polon. B
32, 3605, available at http://th-www.if.uj.edu.pl/acta/

\noindent Klu{\'z}niak W., and Abramowicz M.A., 2002, submitted to
Astron. Astrophys., astro-ph/0203314  

\noindent Shakura N.I., and Sunyaev R.A., 1973, Astron. Astrophys., {\bf
24}, 337 

\noindent Strohmayer T.E., 2001, Astrophys. J. Lett., {\bf 552}, L49 

\bye